\documentclass[prd,aps,graphics]{revtex4}
\usepackage{epsfig}

\def\vk{\textbf{k}}

\def\d{{\rm d}}

\newcommand{\mD}{{\cal D}}
\newcommand{\tPhi}{{\tilde{\Phi}}}

\begin{document}

\title{Modulated  fluctuations from  hybrid inflation}

\author{Francis Bernardeau}
\affiliation{Service de Physique Th{\'e}orique,
             CEA/DSM/SPhT, Unit{\'e} de recherche associ{\'e}e au CNRS,
             CEA/Saclay F-91191 Gif-sur-Yvette cedex, France,}
\email{fbernard@spht.saclay.cea.fr}

\author{Lev Kofman}
\affiliation{CITA, University of Toronto, Toronto,
             Ontario, M5S 3H8, Canada,}
\email{kofman@cita.utoronto.ca}

\author{Jean--Philippe Uzan}
\affiliation{Laboratoire de Physique Th{\'e}orique, CNRS-UMR~8627,
             B{\^a}t. 210, Universit{\'e} Paris XI, F-91405 Orsay cedex, France,\\
             and \\
             Institut d'Astrophysique de Paris, GReCO,
             CNRS-FRE 2435, 98 bis, Bd Arago, 75014 Paris, France.}
\email{uzan@th.u-psud.fr}
\date{\today}
{\bf Preprint:} CITA-2004-11, SACLAY-t04/032

{\bf PACS: 98.70.Vc, 98.80.-k, 02.04.Pc}

%%%%%%%%%%%%%%%%%%%%%%%%%%%%%%%%%%%%%%%%%%%%%%%%%%%%%%%%%%%%%%%%%%%%%%%
\begin{abstract}
Inflation universally produces classical almost scale free Gaussian
inhomogeneities of any light scalars. Assuming the coupling constants
at the time of inflation depend on some light moduli fields, we
encounter the generation of modulated cosmological fluctuations from
(p)reheating. This is an alternative mechanism to generate observable
(almost) scale free adiabatic metric perturbations. We extend this
idea to the class of hybrid inflation, where the bifurcation value of
the inflaton is modulated by the spatial inhomogeneities of the
couplings. As a result, the symmetry breaking after inflation occurs
not simultaneously in space but with the time laps in different Hubble
patches inherited from the long-wavelength moduli inhomogeneities. To
calculate modulated fluctuations we introduce techniques of general
relativistic matching conditions for metric perturbations at the time
hypersurface where the equation of state after inflation undergoes a
jump, without evoking the detailed microscopic physics, as far as it
justifies the jump. We apply this theory to the modulated fluctuations
from the hybrid and chaotic inflations. We discuss what distinguish
the modulated from the inflation-driven fluctuations, in particular,
their spectral index, modification of the consistency relation and the
issue of weak non-Gaussianity.
\end{abstract}
\pacs{{ \bf PACS numbers:} }

\vskip2pc
\maketitle
\vskip 0.15cm
%%%%%%%%%%%%%%%%%%%%%%%%%%%%%%%%%%%%%%%%%%%%%%%%%%%%%%%%%%%%%%%%%%%%%%%

\section{Introduction}\label{sec_intro}

One of the  generic predictions of inflation lies in the fact that vacuum
fluctuations of all light scalar fields, $\chi_a$, minimally
coupled to gravity and with a mass smaller than the Hubble
parameter ($m \ll H$), are universally unstable and appear after
inflation as classical random a priori Gaussian inhomogeneities
with (almost) scale free spectrum ${H}/{\sqrt{2k^3}}$.  The
wavelengths of the fluctuations, $\delta \chi_a(t, \vec x)$, of
such a light scalar field are stretched by inflation and exceed
the Hubble patch after inflation.

One can relate $\delta \chi_a(t, \vec x)$ to the cosmological
scalar metric perturbations in different ways, depending on the
composition of the underlying theory. Indeed, the simplest and
most studied possibility is to assume that there is a single light
scalar field that is the inflaton itself, $\varphi$. The inflaton
fluctuations $\delta \varphi$ are transferred to the scalar metric
perturbations through gravitational interaction~\cite{scalar}.

On general grounds, we may however expect many scalar fields
playing roles during inflation.  There is a broad range of
multiple fields inflationary models with different motivations
behind them, like e.g. double inflation~\cite{KLS85} or hybrid
inflation~\cite{hybr}, where different fields dominate at
different stages of the cosmological evolution.  Multiple fields
were also evoked to design departures from the standard
inflationary predictions: existence of isocurvature
modes~\cite{hybr,iso}, non-scale free spectrum of primordial
fluctuations~\cite{non-flat} or deviation from
Gaussianity~\cite{ng,bu}. In these cases some fields
non-necessarily give dominant contribution to the background
evolution, but during some time give dominant contribution to the
perturbations.  Another corner of the multiple field parameter
space is related to the curvaton scenario~\cite{curvaton}. There,
a newly introduced scalar field, the curvaton, that is indeed not
the inflaton, should be light during inflation, plus dominate
after inflaton decay, plus decay after inflaton but prior BBN,
plus gives a dominant contribution to the metric fluctuations.

A new and more economic idea to generate cosmological perturbations
from modulated fluctuations of couplings was proposed
recently~\cite{modul0,modul1,modul2}.  In the context of multiple
scalar fields theories, it is assumed in these models that some of the
light fields never give dominant contribution neither into background
nor in perturbations during inflation, but contribute to the coupling
constants
\begin{equation}\label{coupl}
\alpha=\alpha(\chi_a) \ .
\end{equation}
Indeed, in string theory~\cite{strings} couplings are in fact the VEV
of moduli fields and, in SUSY theories couplings can also depend on
scalars. As a result, fluctuations of the moduli $\delta \chi_a$
generated during inflation, will manifest themselves in spatial
inhomogeneities of the couplings
\begin{equation}\label{coupl1}
\delta \alpha= \frac{\partial \alpha}{\partial \chi_a} \, \delta \chi_a \ .
\end{equation}
The interaction is not as important during inflation as it is
after inflation. Consider for example simple chaotic inflation.
The background scalar field rolls towards the minimum of the
potential, where it begins to oscillate. Due to the coupling to
the other fields the inflaton decays into radiation in the process
of (p)reheating~\cite{KLS94}. Because of the large scale coupling
inhomogeneities, Eq.~(\ref{coupl1}), at scales much larger than
the causal horizon after inflation, transition from the matter
dominated regime of inflaton oscillations to radiation occurs in
different causal patches not simultaneously, which leads to small
adiabatic metric perturbations after (p)reheating.  After the
moduli $\chi_a$ get pinned down to their minima, the spatial
variations of coupling constants in the late time universe will be
erased.  However, the large scale metric fluctuations that are
produced due to interactions survive as a memory of the primordial
moduli inhomogeneities.

It is actually not mandatory for scalar fields to be light during
inflation. In fact, in the context of $N=1$ supergravity with the
minimal K{{{\"a}}}hler potential during inflation the scalars
typically acquire the mass $m_a \sim H$. Cosmological fluctuations
neither in inflaton sector nor in moduli sector are produced
unless a special care is taken to make at least some of them
light.  This, together with the options to built up inflationary
models, motivate the study of different mechanisms of generation
of cosmological perturbations.

In this article, we investigate this mechanism of {\it modulated
cosmological fluctuations}.  Original papers on the modulated
fluctuations \cite{modul1,modul2} discussed metric perturbations
generated from the decay of inflaton oscillations.  However, there is
another important class of hybrid inflationary models, which typically
emerges in supergravity and the string theory/brane cosmology.  The
two field $(\varphi, \sigma)$ hybrid inflation has the effective
potential
\begin{equation}\label{veff}
 V_{\rm eff} = \frac{\lambda}{4}\left(\sigma^2 - v^2\right)^2 +
 \frac{1}{2}g^2\varphi^2\sigma^2 + V(\varphi)
\end{equation}
where $V(\varphi)$ is the inflaton potential.  This potential
contains two couplings $\lambda$ and $g^2$, which define the
end-point of inflation and the field dynamics.  Assuming these
couplings are moduli dependent, we encounter modulated
fluctuations in the hybrid inflation.  This is the main novel idea
of the paper which we describe and develop in
Section~\ref{sec_hybrid}.

One may think of different aspects of the theory of modulated
fluctuations, related to the nature of the moduli and
dependencies, Eq.~(\ref{coupl}), details of preheating,
thermalization, moduli evolution etc.  Here we concentrate on the
cosmological General Relativistic part of the story, namely, how
to derive cosmological metric perturbations from the coupling
perturbations, Eq.~(\ref{coupl1}). In the chaotic inflation
typically inflaton oscillations decay through the non-perturbative
effect of parametric resonance of particle creation~\cite{KLS94}.
This leads to copious production of particles in
out-of-equilibrium state. Further interactions between particles
relax them towards thermal equilibrium. Perturbative regime of
particle interactions takes place at the latest stages of
transition from inflation towards hot
radiation~\cite{Felder:2000hr}. In hybrid inflation, preheating
after inflation has the character of the tachyonic
preheating~\cite{tachprech}, accompanying the symmetry breaking
after inflation~\cite{tachprech2}. Tachyonic preheating leads to
generation of particles out-of-equilibrium with subsequent
thermalization of them. Coupling constants of interaction appears
at different stages of preheating and thermalization. However, for
the production of modulated cosmological perturbations it is
essential that couplings are responsible for the change of the
equation of state.  If the equation of state is not changing,
modulated perturbations are not generated.  In this respect we
shall concentrate at the first instance where equation of state
after inflation is changing due to the couplings. In the case of
chaotic inflation it happens when an effective matter equation of
state of the coherent inflaton oscillations is replaced by the
radiation in the very fast process of preheating.  In the case of
hybrid inflation, the vacuum-like (inflation) equation of state is
replaced by radiation in the very quick process of tachyonic
preheating.

So far the formalism of the modulated cosmological fluctuations
was considered for the toy model of slow perturbative reheating
\cite{modul1,modul2,mata,vernizzi}, often with the Yukawa type
interaction $\phi \bar \psi \psi$ between inflaton and fermions.
 This is fair enough to see in principle
that metric fluctuations are generated from the moduli
inhomogeneities. However, the modern theory of the transition
between inflation and radiation is described by the theory of
preheating. Even for case of  fermions their production is
non-perturbative and significantly modified by preheating
\cite{fermion}.  In this paper, on the technical side, we suggest
the method to treat generation of modulated cosmological
fluctuations in the inflationary models with preheating. We will
use the fact that in all cases preheating is very short, and from
the general relativistic point of view can be considered as the
instant jump of the equation of state. It is then convenient to
use the formalism of matching conditions of the geometrical
quantities at the time of the transition~\cite{Deruelle:1995kd}.
For this, we do not need to know the microscopic details of
preheating and thermalization.

In Section~\ref{sec_hybrid}, we first discuss the basics of the
modulation of the couplings and some model building aspects that
have to be fulfilled for such a mechanism to be efficient. After
recalling the basics of the matching conditions in
Section~\ref{Sec-match}, we then apply them to the standard single
field inflationary case in Section~\ref{sec_infl}.  We then turn
to the case of modulated fluctuations in
Section~\ref{sec_modfluc}. We will discuss the main features of
the mechanism in Section~\ref{sec_concl} and in particular
emphasize that it allows to extend the standard consistency
relation of inflation.

\section{Modulated fluctuations of couplings
 in hybrid inflation}\label{sec_hybrid}

Consider a model of hybrid inflation. The basic shape of the effective
potential is given by Eq.~(\ref{veff}), where
$\varphi$ is the inflaton $\sigma$ is another scalar field, which is
massive during inflation but whose effective mass changes sign at the
critical value of the inflaton value
\begin{equation}\label{phic}
  \varphi_c=\frac{\sqrt{\lambda}\,v}{g}.
\end{equation}
The point where $\varphi=\varphi_c,\sigma=0$ is a bifurcation
point. For $\varphi > \varphi_c$ the squares of the effective masses
of both fields are positive and the potential has a minimum at
$\sigma=0$. For $\varphi < \varphi_c$ the potential has a maximum at
$\sigma=0$. The global minimum is located at $\varphi=0$ and
$|\sigma|= v$. However, at $\varphi > \varphi_c$ the effective
potential has a valley along $\sigma = 0$. In this model, inflation
occurs while the $\varphi$ field rolls slowly in this valley from
large values towards the bifurcation point.  At bifurcation point the
symmetry breaking occurs.  Recall, however, that immediately after
the bifurcation point the field $\sigma$ has a negative mass square.
Hence, dynamics of the symmetry breaking is accompanying by the
tachyonic instability of inhomogeneous modes.  It results in the very
rapid decay of the homogeneous fields into inhomogeneous modes in the
non-linear regime of tachyonic preheating \cite{tachprech,tachprech2}
much before the global potential minimum is reached. For what we are
interested in this bath of inhomogeneous modes essentially behaves
like a radiation fluid.  In the following we will simply assume the
transition between inflation and radiation domination to be
instantaneous

In general in hybrid models $\varphi$ and $\sigma$ can be viewed
as scalar fields coming from a much larger scalar sector of the
theory.  We then assume that there exist a set of light scalar
fields, $\chi_a$ and that the couplings $\lambda$ and $g$ absorb
the dependence on these scalars as
\begin{equation}
 \lambda=\lambda(\chi_a),\quad
  g = g(\chi_a)\quad
 \Longrightarrow\quad
 \varphi_c = \varphi_c(\chi_a) \ .
\end{equation}
 $\lambda$, $g^2$ are determined by the VEVs of these
fields. Since the fields $\chi_a$ are light, they develop
super-Hubble inhomogeneities  so that the symmetry breaking that
terminates the inflationary stage does not occur at the same time
everywhere and is modulated over space. The fluctuations of the
light fields modulate the couplings $\lambda$ and $g^2$ which then
have spatial fluctuations in a way that depends on their
dependence on the $\chi_a$ VEVs,
\begin{eqnarray}\label{6}
  \delta\lambda &\approx& \sum_a \frac{\partial \lambda}{\partial
  \chi_a}\delta \chi_a;\\
  \delta g^2 &\approx& \sum_a \frac{\partial g^2}{\partial
  \chi_a}\delta \chi_a.
\end{eqnarray}
It follows that our
model consists of the inflaton, $\varphi$, a Higgs-like field,
$\sigma$ and the light fields, $\chi_a$ with a potential of the form
\begin{equation}\label{veff2}
 V_{\rm eff} = \frac{1}{4}\lambda(\chi_a)\left(\sigma^2 - v^2\right)^2 +
 \frac{1}{2}g^2(\chi_a)\varphi^2\sigma^2 + V(\varphi) \ .
\end{equation}
In principle, moduli may have potentials $U(\chi_a)$, which for
simplicity are assumed to be are negligible during inflation, or
included into (\ref{veff2}), as we will see below.

In Eq.~(\ref{veff2}) the dependencies of the coupling constant on the
moduli fields $\chi_a$ have quite different status.
Let us inspect the conditions in the theory  (\ref{veff2})
 which would keep the moduli light
during inflation. When
 $\varphi>\varphi_c$ and $\sigma=0$  the
 equations for the fields $\chi_a$ are
\begin{equation}\label{field}
 \ddot\chi_a + 3H\dot\chi_a  +
 \frac{v^4}{4}\frac{\partial \lambda}{\partial \chi_a}=0  \ .
\end{equation}
Equations for its fluctuations $\delta \chi_a \hbox{e}^{i \vec k.\vec x}$
are
\begin{equation}\label{fluc}
\delta\ddot\chi_a + 3H\delta\dot\chi_a +
 \frac{k^2}{a^2}\delta\chi_a +
 \frac{v^4}{4}\frac{\partial^2
 \lambda}{\partial \chi_a \partial \chi_b} \, \delta\chi_b=-
 \frac{v^4}{2}\frac{\partial \lambda}{\partial \chi_a} \Phi
 + 4\dot\chi_a\dot\Phi \ .
\end{equation}
During
the inflationary stage $\varphi>\varphi_c$ the effective potential
in which $\chi_a$ evolves is $\lambda(\chi_a)\,v^4/4$

Suppose the background value of $\chi_a$ is order of $M$
and natural argument of couping is ${\chi_a}/{M}$.
Hence, the fluctuations of the  fields $\chi_a$ have a  mass
$m_a^2=({v^4}/{4}){\partial \lambda}/{\partial \chi_a}
 \sim H^2{M_p^2}/{M^2}$, where $M_p$ is the Planck mass.
 But as we will see
in the next section, the right amplitude of modulated fluctuations
is achievable if the mass scale $M$ is of order or smaller than
$M_p$. Thus,  $\delta \chi_a$ are heavy unless
the coupling $\lambda$ has no dependence on moduli.
There is no such reservation for $g^2$. In the
following we will see that supergravity D-term inflation precisely
provides us with such a pattern for the moduli dependence.

When the mass of $\chi_a$ is small, and its contribution to the
background geometry is negligible, its kinetic energy is also
small, e.g.  $\dot\chi_a \approx 0$.  In this case the
fluctuations $\delta \chi_a$ can be considered as the light test
field at given background driven by inflaton $\varphi$. It is easy
to see that $\chi_a$ also do not influence the evolution of the
inflaton field and fluctuations during inflation. Indeed we have
equations for inflaton fluctuations $\delta \phi$ and metric
fluctuations $\Phi$ (in the longitudinal gauge) during this phase
\begin{eqnarray}
 &&\dot\Phi + H\Phi = \frac{1}{2M_p}\left(\dot\varphi\delta\varphi
   +\dot\chi_a\delta\chi_a\right),\label{pert1}\\
 &&\delta\ddot\varphi + 3H\delta\dot\varphi +
 \frac{k^2}{a^2}\delta\varphi = -2V'\Phi + 4\dot\varphi \dot\Phi
 -V''\delta\varphi.\label{pert2}
 \label{pert3}
\end{eqnarray}
It follows that the fluctuations of the light fields $\chi_a$ do not
contribute to the metric fluctuations in Eq.~(\ref{pert1}) during
inflation, so that the set of Eqs.~(\ref{pert1}-\ref{pert2}) reduces
to the one of the inflaton coupled to metric perturbation, as in
single field inflation.

To summarize, since $\sigma=0$ during inflation, the function
$g(\chi_a)$ does not enter the perturbation equations during
inflation so that the light fields can be considered as test
fields. As anticipated, the light fields influence only the end of
inflation by modulating over space the time at which the symmetry
breaking occurs, imprinted in the hypersurface $\varphi =
\varphi_c$. They are subdominant with respect to the VEVs of
$\varphi$ and $\sigma$ that drive the evolution of the background
and do not affect the generation of metric perturbations during
inflation.  We expect the cumulative fluctuations after inflation
to inherit both from the metric fluctuation during inflation and
from the modulated transition, hence opening the possibility of
non trivial phenomenological consequences.

\section{Moduli in  couplings of  Sugra D-term inflation}\label{SugraModel}

In this section we slightly step aside of the main topic of the paper,
modulated cosmological fluctuations in the generic  hybrid inflation.
We consider a specific example of the 
sugra D-term inflation, to illustrate how the
dependence of couplings on moduli can be originated.
Although in general, D-term inflation is similar to the hybrid inflation, there are
some differences. In particular, 
slow roll regime of
 inflation may be ended even before
$\phi$ reaches the bifurcation point \cite{BD,LR}.
  D-term inflation
is reduced to the hybrid inflation  in the limit of small $g^2$.
All our results are immediately applicable for this limit of
 D-term inflation. In more general case of  D-term inflation
the equation of stata may be changed twice, at $\phi_c$ and even before 
 at some  hypersurface $\phi_e > \phi_c$,
but  $\phi_e$ is still  spatially varying due to the moduli dependence.
As a result the magnitude of modulated fluctuations may be even greater
than our estimations below (where only a single jump in the equation of state
is considered). 
We notice but do  not consider these effects in the paper.

Let us recall how potentials of the form of Eq.~(\ref{veff}) can be
obtained from particle physics models. Such potentials are indeed
prototypes of models of inflation motivated by supergravity (including
low-energy string theory) such as $F-$ or $D-$term inflations (see
e.g. Ref.~\cite{susyreview} for a review). It might be worth having in
mind that the more generic $P-$term inflation has both these models as
limiting cases \cite{Pterm}. Note also that brane-antibrane
$D-\overline{D}$ systems in superstring theory produce 4-dimensional
effective potentials like (\ref{veff}).

In $F$-term inflation we always have $g^2=\lambda$ and no
modulated fluctuations can be generated. We will not consider this
case any further. On the contrary a $D-$term inflation driven by a
non-zero Fayet-Ilioupoulos $D$-term does not lead to any specific
relation between $g^2$ and $\lambda$.

For further discussion recall that a generic $N=1$ supergravity
lagrangian including interaction with matter and Yang-Mills fields
in $3+1$ dimensions is built from three arbitrary functions, the
K{{{\"a}}}hler potential $K(\chi_a^*, \chi_a)$ which encodes the
kinetic term of the scalar fields, the superpotential $W(\chi_a)$
and the kinetic terms $f_{\alpha \beta}(\chi_a)$ for the vector
multiplet fields, $\left[{\rm Re} f_{\alpha \beta}(\chi_a) \right]
\, F^{\alpha}_{\mu\nu} F^{ \beta \mu\nu}$. We will use notation
and the form of the supergravity lagrangian, see Eq. (5.15) of
Ref.~\cite{sugra}, adapted to cosmology.

The simplest model of D-term hybrid inflation~\cite{BD} consists
of three (left) chiral superfields $\Phi_i$: the inflaton and two
fields of opposite charge under a local $U(1)$. The potential for
this model comes from the superpotential
\begin{equation}\label{w}
W = \sqrt2 g\, \Phi \Phi_+ \Phi_- \ ,
\end{equation}
and the D-term
\begin{equation}\label{d}
D = {\sqrt{\lambda} \over 2} \left(2|\Phi_+|^2 - 2|\Phi_-|^2 -
v^2 \right) \ ,
\end{equation}
where the fields $\Sigma_\pm$ have charges $\pm \sqrt{\lambda}$ and
the Fayet-Illiopolous term is $\sqrt{\lambda}v^2/2$.  We may choose
the fields to be real and define $\sigma_\pm \equiv
\sqrt{2}|\Phi_\pm|$ and $\phi \equiv \sqrt{2}|\Phi|$, allowing for
canonical kinetic terms. In terms of these fields, in the global SUSY
limit the potential reduces to the form
\begin{equation}\label{potd}
V=V_{\rm D}+V_{\rm F}   = {\lambda\over4}\left(\sigma_+^2 - \sigma_-^2 - v^2\right)^2 +
{g^2\over2} \left(\phi^2\sigma_+^2 + \phi^2\sigma_-^2 +
\sigma_+^2 \sigma_-^2 \right) \ .
\end{equation}
When $\phi$, which plays the role of the inflaton, is large enough
both $\sigma_+$ and $\sigma_-$ have a large positive mass and are
forced to be zero. Supersymmetry is however broken and the 1-loop
corrections give an extra term in the potential, which is only
$\phi$ dependent so that it rolls down toward $\phi=0$. This model
exactly matches Eq. (\ref{veff}) noting that during the whole
evolution $\sigma_-$ is forced to be zero.

In more general context of supergravity, $D-$ term has pre-factor
$\left[{\rm Re} f_{\alpha\beta}(\chi_a) \right]^{-1}$ absorbed in
the coupling $\lambda$, $\lambda(\chi_a) \to \lambda \times[{\rm
Re} f(\chi_a)]^{-1}$.  Thus, a large mass of $\chi_a$ can be
avoided by simply assuming that ${\rm Re} f_{\alpha \beta}=1$. The
effective coupling $g^2$ can then be made dependent on $\chi_a$
through the K{{{\"a}}}hler potential.

We give a toy model example where we want to observe that the moduli
dependence can appear in $g^2$ while do not appear in $\lambda$. The
K{{{\"a}}}hler potential intervenes in the $F-$term part of the
potential
\begin{equation}\label{FtermSugra}
  V_F=e^{K/M_P^2}\left[K^{j^*i} \mD_{j^*} W  \, \mD_{i}W
-3\,\frac{W^2}{M_P^2}
  \right],
\end{equation}
where $\mD_i={1/M_P^2}\, {\partial K}/{\partial \Phi_i}+{\partial
}/{\partial \Phi_i}$, $K^{j^*i}$ is the inverse matrix to
$K_{ij^*}\equiv
\partial^2 K/\partial \Phi_i\partial\Phi_j^* $. The $D$ term on
the other hand is given by
\begin{equation}\label{DExpSugra}
  D=\frac{\sqrt{\lambda}}{2}\left(q_i\,K_i\phi_i-v^2\right)
\end{equation}
where $K_{i}\equiv
\partial K/\partial \Phi_i$, $q_i$ is the charge of corresponding scalar. 

For a standard simple string theory toroidal compactification
scheme~\cite{witten168} for which,
\begin{equation}\label{KExp1}
  K=-3\log \left(t+t^*-\sum_i\vert\Phi_i\vert^2\right),
\end{equation}
and in case the VEVs of the fields (and in particular that of
$\Phi_0$ which is non vanishing during inflation) are small
compared to $\vert t\vert$ the computations can be easily
completed. We have $K_i\approx 6\phi_i^*/(t+t^*)$ and
$K_{ij^*}\approx 6/(t+t^*)\delta_{ij}$. This latter matrix can
easily be inverted. As a result the fields $\tilde\Phi_i\equiv
\sqrt{3/(t+t^*)}\,\Phi_i$ are the fields that have a standard
kinetic term in the low energy limit. We see that in this case the
$t$ dependence in Eq. (\ref{DExpSugra}) drops. Unfortunately this
is also the case in the expression of $V_F$.

If however the compactification is made in such a way that there
are 3 different moduli directions $t_i$ then the expression of
$V_F$ can be made dependent on the moduli values. This is the case
for instance if,
\begin{equation}\label{Kpot}
  K=-\log \left(t_1+t_1^*-\sum_i\vert\Phi_i\vert^2\right)
  -\log(t_2+t_2^*)-\log(t_3+t_3^*).
\end{equation}
In this case $K_i=2\Phi_i^*/(t_1+t_1^*)$ and
$K_{ij^*}=2/(t_1+t_1^*)\delta_{ij}$. The kinetically regularized
fields are $\tilde\Phi_i\equiv \sqrt{1/(t_1+t_1^*)}\,\Phi_i$ from
which we get,
\begin{eqnarray}\label{FDterm2}
  V_F&=& {\tilde g}^2 \left[\vert\tPhi\tPhi_+\vert^2+\vert\tPhi\tPhi_-
  \vert^2+\vert\tPhi_-\tPhi_+\vert^2\right] \\ \nonumber
\tilde g^2&=&\frac{(t_1+t_1^*)^2}{(t_2+t_2^*)(t_3+t_3^*)}
  g^2 \ ,
\end{eqnarray}
which indeed leads to a moduli dependent effective $g^2$. Note that if
each $\Phi$ is associated with a different moduli direction the
dependence in $g^2$ also vanishes. We see that unless the K{\"a}hler
potential has very specific features, the effective coupling constant
$g^2$ is dependent on the moduli fields. This is what we were looking
for. The picture we obtained here is not as simple as the one
described in the introduction since it implies that the inflaton
potential, $V(\phi)$ here, coming from the radiative correction to the
potential induced by the breaking of the boson-fermion mass
equalities, depends on $g$, therefore on the moduli fields. It means
that the inflaton is actually a combination of $\Phi$ and of the
moduli.  The picture sketched in the beginning will be recovered if
the dependence with the latter is small enough. Note also that anyhow
the theory we present is incomplete because it does not provide for a
stabilization mechanism for the VEVs of the moduli.
The  problem of stabilization in the context of D-term inflation is discussed e.g.
in \cite{ren}, see refs therein.
  We will then not
exploit any further this specific model. We value it however because
it shows that modulated inflation should be rather generic is
realistic models of Sugra.

In the following sections we derive the geometrical theory of
modulated cosmological fluctuations which is applicable for the hybrid
inflation, as well as from the chaotic inflation.

\section{Junction conditions for metric perturbations
at the time hypersurface}\label{Sec-match}

In this and the following sections we turn to technical part of the
paper.  We will deal with the general relativistic aspects of the
cosmological metric fluctuations. First, in this section we will
remind the general formalism of the matching
conditions~\cite{matching} of two geometries on the different
sides (past and future) of a time hypersurface. This spacelike
hypersurface, $\Sigma$, also divides the matter contents in the
universe, which has different equations of state on different
sides of $\Sigma$. In subsequent sections, we apply this general
formalism to different situations.

\begin{figure}[b]
 \centerline{\epsfig{figure=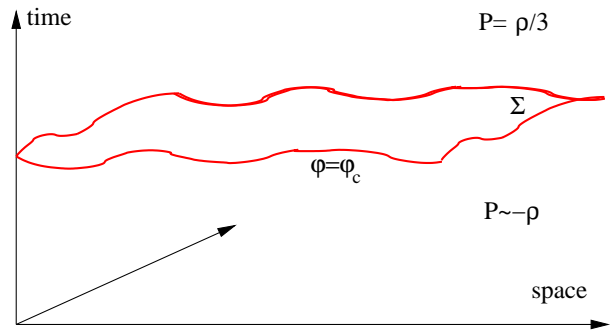,width=8cm}}
 \caption{Inflationary period is matched to a radiation dominated era
 on a spacelike hypersurface $\Sigma$ defined by $q(\varphi,\chi_a)
 \equiv\varphi-\varphi\c(\chi_a)=0$.}
 \label{plot}
\end{figure}

We begin with the derivation of the matching conditions in the
cosmological context. We assume that the transition between two eras,
e.g. inflation and radiation dominated, takes place on a three
dimensional spacelike hypersurface defined by
$$
\Sigma=\{q = {\rm constant}\},
$$
where $q$ is a scalar to be specified,
see Figure~\ref{plot}.
 We focus on scalar and
tensor modes of metric perturbations
 and assume that the spatial sections of the universe
are flat and work in longitudinal gauge. It follows that the
metric of spacetime takes the form
\begin{equation}\label{a1}
 \d s^2_\pm=a^2_\pm(\eta_\pm)\left[-(1+2\Phi_\pm)\d\eta^2_\pm +
            \left\{(1-2\Phi_\pm)\delta_{ij}+h_{ij}^\pm\right\}
            \d x^i\d x^j\right]
\end{equation}
where the index $-$ and $+$ refer respectively to the two eras,
before and after $\Sigma$.
$h_{ij}$ is a symmetric traceless ($h_i^i=0$) transverse
($\partial_i h^{ij}=0$) tensor describing the gravitational waves.
Note that the  conformal
 times $\eta_{\pm}$ in both era are a priori different. We
also split $q$ as
\begin{equation}\label{a2}
  q =\bar q + \delta q,
\end{equation}
$\delta q$ being the perturbation in longitudinal  gauge.

The junction conditions~\cite{matching} reduce to the continutity of
the induced three dimensional metric on $\Sigma$ and of the extrinsic
curvature of $\Sigma$, see
Ref.~\cite{Deruelle:1995kd,ds,Deruelle:1997}. The normal unit vector to
the hypersurface $\Sigma$ is given by
\begin{equation}\label{a3}
 n_\mu=\frac{\partial_\mu q}{\sqrt{-\partial_\alpha q
 \partial^\alpha q}}
\end{equation}
so that the induced three dimensional metric on $\Sigma$ takes the form
\begin{equation}\label{a4}
 \gamma_b^a= a^2\left[
  \left(1-2(\Phi+{\cal H}\frac{\delta q}{q'}\right)
  \delta_b^a + h^{a}_b
 \right]
\end{equation}
and its extrinsic curvature
\begin{equation}\label{a5}
 K_b^a=\frac{1}{a}\left[-{\cal H}\delta_b^a
      + \left({\cal H}\Phi + \Phi' +({\cal H}'-{\cal H}^2)
      \frac{\delta q}{q'}\right)\delta_b^a
      +\partial^a\partial_b \frac{\delta q}{q'}
      + \frac{1}{2}h^{\prime a}_b
 \right]
\end{equation}
where $a,b$ run from 1 to 3, and prime is the derivative
wrt $\eta$, ${\cal H}=\frac{a'}{a}$.
 It follows that the matching
conditions for the background geometry
reduce to
\begin{equation}\label{a6}
 \left[a\right]_\pm=0,\qquad
 \left[{\cal H}\right]_\pm=0
\end{equation}
 where $[X]_\pm\equiv X_+ -X_-$.
Another words, the scalar factor and its time derivative
are continues through $\Sigma$.

Matching conditions for perturbations are
 split into
\begin{equation}\label{a7}
 \left[\Phi\right]_\pm=0,\qquad
 \left[\Phi' +{\cal H}'\frac{\delta q}{q'}\right]_\pm=0,\qquad
 \left[\frac{\delta q}{q'}\right]_\pm=0,
\end{equation}
for the scalar perturbations and 
\begin{equation}\label{a8}
 \left[h_{ij}\right]_\pm=0,\qquad
 \left[h'_{ij}\right]_\pm=0
\end{equation}
for the gravitational waves. Equations (\ref{a7}) and (\ref{a8}) are
the basis for the applications below.

\section{Junction conditions for inflaton
 driven metric fluctuations}\label{sec_infl}

In order to present the method and the notation, we re-derive the
standard result for the scalar metric fluctuations driven by inflaton
fluctuations, using the general formalism of junction conditions for
metric fluctuation outlined above (see also
Refs.~\cite{Deruelle:1995kd,ds}).

\subsection{Long wavelength  modes evolution}

So for the time being let us assume that inflation is driven by a
slow-rolling scalar field $\varphi$, we can introduce the slow
roll parameters by
\begin{equation}\label{srpara}
 \varepsilon = \frac{M_p^2}{16\pi}\left(\frac{V'}{V}\right)^2,\qquad
 \eta =  \frac{M_p^2}{8\pi}\left(\frac{V''}{V}\right) \ .
\end{equation}
 Solving for
$\varepsilon=1-{\cal H}'/{\cal H}^2$, gives the expression of the
 parameter
\begin{equation}\label{sr4}
 {\cal H}_-= \frac{1+\varepsilon}{-\eta_-} \ ,
\end{equation}
where conformal time during inflation is $\eta_-$. It is convenient to
introduce the two quantities defined by
\begin{equation}\label{b1}
 {\cal R}=\Phi+\frac{2}{3}\frac{\Phi'+{\cal H}\Phi}{{\cal
 H}(1+w)},\qquad
 \zeta= \Phi + {\cal H}\frac{\delta\rho}{\rho'},
\end{equation}
that correspond to the curvature perturbation in flat slicing gauge
and in comoving gauge. For a general perfect fluid, including the
particular case of a single scalar field, the perturbation equations
take the form~\cite{mfb}
\begin{eqnarray}
 &&\Delta\Phi = \frac{\kappa a^2}{2}\left[\delta\rho - 3{\cal
 H}\rho(1+w)v\right]\label{b2}\\
 && \Phi' + {\cal H}\Phi = -\frac{\kappa a^2}{2}\rho(1+w)v\label{b3}
\end{eqnarray}
where $\delta\rho$ and $v$ are respectively the density
perturbation and the velocity perturbation in Newtonian gauge and
with $\kappa\equiv8\pi G$. It follows that the density fluctuation
can be expressed as,
\begin{equation}\label{b4}
\frac{\kappa a^2}{2}\delta\rho= \Delta\Phi + 3{\cal H}\left(\Phi'
+ {\cal H}\Phi\right).
\end{equation}
and that the two quantities defined in Eq.~(\ref{b1}) are related,
using that ${\cal H}'-{\cal H}^2={\cal H}\rho'/\rho$, by
\begin{equation}\label{b6}
 {\cal R} +\frac{1}{3}\frac{\Delta\Phi}{{\cal H}'-{\cal H}^2} =
 \zeta
\end{equation}
so that they are equal for super-Hubble modes. The evolution equation
for the gravitational potential, see Ref.~\cite{mfb}, can be shown to imply
\begin{equation}\label{b7}
{\cal R}'=0,\qquad \zeta'=0,
\end{equation}
for super-Hubble modes. In fact, the solution evolution equation of
the gravitational potential takes the general form
\begin{equation}\label{b9}
 \Phi=\frac{{\cal H}}{a^2}\left(B + A\int(1+w)a^2\d\eta\right).
\end{equation}
The coefficient $A$ corresponds to the growing mode and $B$ to a
decaying mode. If the equation of state vary continuously from
$w=-1+2\varepsilon/3$ during inflation to $w=1/3$ during the radiation
era, we obtain that the gravitational potential in RDU, after the
decaying modes have become negligible,
\begin{equation}\label{b66}
 \Phi\sim \frac{2}{3}\frac{1+\varepsilon}{\varepsilon}A
\end{equation}
where $A$ is fixed during inflation. Since ${\cal
R}\simeq\Phi/\varepsilon\simeq A/\varepsilon$ during inflation and
${\cal R}\simeq 3\Phi/2$ during RDU, the solution of Eq.~(\ref{b9}) is
equivalent to the continuity of ${\cal R}$, Eq.~(\ref{b7}). Note also
that Eq.~(\ref{b9}) implies that the variation of the gravitational
potential when the scale factor changes behavior changes from
$a\propto t^{p_1}$ to $a\propto t^{p_2}$ at $t=t_*$ is $\Phi(t\gg
t_*)/\Phi(t<t_*)=(1+p_1)/(1+p_2)$ (see Refs.~\cite{Deruelle:1995kd}).\\

\subsection{Derivation by means of the junction conditions}

Let us now do the same exercise but by means of the junction
conditions and assume inflation suddenly ends with a transition to
a radiation era. As long as the background dynamics is concerned,
the matching conditions for the background quantities, that is the
continuity of the scale factor and of the Hubble parameter,
Eq.~(\ref{a6}), imply that the transition happens at
$\eta_+=\eta_*=-(1-\varepsilon)\eta_-$ and we have
\begin{equation}\label{bgd}
a_-(\eta_-)=C/(-\eta_-)^{1+\varepsilon},\qquad
a_+(\eta_+)=C/[(1+\varepsilon)\eta^{1+\varepsilon}_*](\eta_+/\eta_*).
\end{equation}
Now, the transition is due to a sudden change in the equation of state
and takes place on a constant density
hypersurface~\cite{Deruelle:1995kd}. The matching conditions (\ref{a7}) imply
that
\begin{equation}
\left[\frac{\delta\rho}{\rho'}\right]_\pm=0,\qquad
\left[\Phi\right]_\pm=0,\qquad \left[\Phi'+({\cal H}'-{\cal
H}^2)\frac{\delta\rho}{\rho'}\right]_\pm=0.
\end{equation}
The first two conditions and the continuity of ${\cal H}$ imply
that
\begin{equation}\label{zz}
\left[\zeta\right]_\pm=0
\end{equation}
and thus that
\begin{equation}
\left[{\cal R} +\frac{1}{3}\frac{\Delta\Phi}{{\cal H}'-{\cal
H}^2}\right]_\pm=0.
\end{equation}
The first condition implies that
$$
\left[\frac{\Delta\Phi-3{\cal H}(\Phi'+{\cal
H}\Phi)}{1+w}\right]_\pm=0
$$
which reduces to
\begin{equation}\label{b8}
\left[\frac{3{\cal H}(\Phi'+{\cal H}\Phi)}{1+w}\right]_\pm=0
\end{equation}
on super-Hubble scales. It follows that the matching conditions
imply the continuity of $\zeta$, Eq.~(\ref{zz}) and the continuity
of ${\cal R}$ on super-Hubble scale. Note however that this
conclusion relies strongly on the fact that $\delta \rho/\rho'$ is
continuous, and thus on the choice of the matching surface (see
Ref.~\cite{ds} for further discussions on this issue). Note that
matching on a constant $\varphi$ hypersurface will have implied,
from Eq.~(\ref{a7}) that $[{\cal R}]_\pm=0$, instead of
Eq.~(\ref{zz}). Interestingly, Eq.~(\ref{b6}) shows that for
supper-Hubble modes, it is equivalent to match on a constant field
or constant density hypersurface.

\begin{figure}
 \centerline{\epsfig{figure=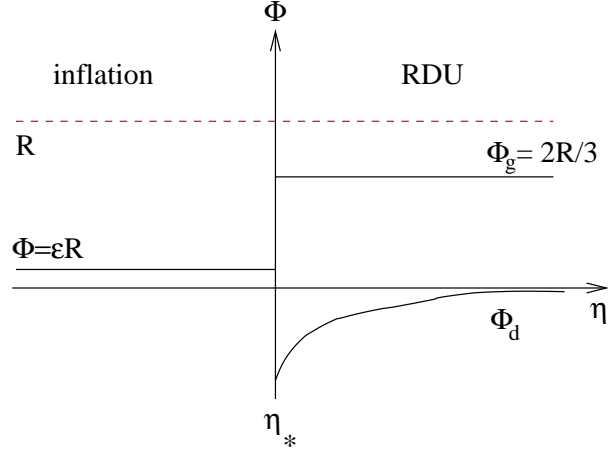,width=8cm}}
 \caption{The evolution of the gravitational potential $\Phi$ through the transition. Both
 ${\cal R}$ and $\Phi$ are continuous if the transition is a constant density hypersurface. $\Phi_g$
 and $\Phi_d$ refer respectively to the growing and decaying modes.}
 \label{fig11}
\end{figure}

The general solution for the gravitational potential, Eq.~(\ref{b9}),
implies that  $\Phi=A_- + B_-(-\eta)$ and $\Phi=A_+ +
B_+/\eta^3$ respectively during inflation and RDU. The continuity
of $\Phi$ and the condition (\ref{b7}) imply that
\begin{equation}\label{zozo}
A_+=\frac{2}{3}\frac{1+\varepsilon}{\varepsilon}A_-,\qquad B_+
=\frac{1}{3}\frac{\varepsilon-2}{\varepsilon}A_+,
\end{equation}
if we neglect the decaying mode during inflation. At a time still
in the RDU but far enough from the transition, we get that
\begin{equation}\label{b10}
 \Phi_+ \sim \frac{2}{3}\frac{1+\varepsilon}{\varepsilon}\Phi_-,
\end{equation}
which is the standard result, Eq.~(\ref{b66}). Again, since ${\cal
R}_-=\Phi_-/\varepsilon$ and ${\cal R}_+ = 3\Phi_+/2$, Eq.~(\ref{b9})
is equivalent to the continuity of ${\cal R}$. The behavior of the
different quantities is depicted on figure~\ref{fig11}. Note that the
growing mode of the gravitational potential inherits a contribution
from the gravitational potential during inflation ($\Phi_-$) and a
contribution form the density perturbation on the matching surface
[$(2-\varepsilon)\Phi_-/3\varepsilon$]. In modulated inflationary
model, this second contribution will be modified, as we will see in
the next section.

\subsection{Initial power spectra and the consistency relation}

The preceding argument simply shows that in the standard case it
is equivalent to use the matching conditions or the continuity of
$\zeta$ to relate the gravitational potential generated during
inflation to the one in the radiation era. Indeed, one still needs
to determine the gravitational potential generated during
inflation that can be obtained from the quantification of the
density fluctuations during inflation. This is part of the
standard lore.

For the scalar modes, let us introduce the Mukhanov
variables~\cite{mfb}
\begin{equation}
 u= z {\cal R} =z \left(\Phi+\frac{{\cal H}}{\varphi'}\delta\varphi\right),\qquad
 z = \frac{a\varphi'}{{\cal H}}
\end{equation}
in terms of which the equation of evolution of the scalar modes is
\begin{equation}\label{smodes}
 u'' +(k^2 - z''/z)u =0.
\end{equation}
In terms of the slow-roll parameter $z''/z=(\nu^2-1/4)/\eta^2$
with $\nu=3/2 + 3\varepsilon -\eta$ so that the general solution
of Eq.~(\ref{smodes}) can be expressed in terms of Hankel
functions. On super-Hubble scales, it reduces to
\begin{equation}
u_k \simeq \frac{1}{\sqrt{2k}}(-k\eta)^{1/2-\nu}.
\end{equation}
We deduce that the curvature perturbation, ${\cal R}_k=u_k /z$, is
given by
\begin{equation}
 {\cal R}_k \simeq \frac{1}{\sqrt{2k^3}}\frac{H}{M_p}\frac{1}{\sqrt{\varepsilon}}
 (-k\eta)^{\eta-3\varepsilon}
\end{equation}
before the transition. Defining the power spectrum of any field
$X_\vk$ as
\begin{equation}\label{defdeP}
 \left<X_\vk
 X_{\vk'}\right>=\frac{2\pi^2}{k^3}P_X(k)\delta(\vk-\vk'),
\end{equation}
one obtains that
\begin{equation}
P_{\cal R}^{1/2}=\frac{1}{2\pi}\left({H\over
M_p}\right){1\over\sqrt{\varepsilon}}
(-k\eta_-)^{\eta-3\varepsilon}\simeq{1\over\varepsilon}P^{1/2}_{\Phi_-}\simeq{2\over
3}P_{\Phi_+}^{1/2},
\end{equation}
where the last equality derives from Eq.~(\ref{b10}).

Concerning the gravitational waves, one can follow the same routes but
now the matching conditions Eq.~(\ref{a8}) are trivial.  Introducing
the variable~\cite{mfb} $ u_T= a M_p h$, the evolution of the tensor
modes is dictated by the equation
\begin{equation}\label{tmodes}
 u_T'' +(k^2 - a''/a)u_T =0.
\end{equation}
The general solution is given in terms of Hankel function and the
growing mode on super-Hubble scales is given by
\begin{equation}\label{tens1}
 h_+\sim h_- \sim \frac{1}{\sqrt{2k^3}}\frac{H}{M_p}(-k\eta)^{-\varepsilon}
\end{equation}
from which one deduces a consistency relation between the relative
amplitude between scalar and tensor contributions and the slow
roll parameter
\begin{equation}\label{consist}
 {T\over S}        = \varepsilon = - n_T/2
\end{equation}
where $T$ and $S$ are measuring the amplitude of the power spectra
of respectively the tensor and scalar curvature modes and where
$n_T$ is the tensor modes spectral index.

\section{The case of modulated fluctuations}\label{sec_modfluc}

Now we turn to the main subject of the paper, namely the generation of
modulated fluctuations.  Contrary to the previous case of inflaton
driven fluctuations, the transition between the inflationary era and
radiation era does not take place on a constant density hypersurface
but on a hypersurface of constant value of the bifurcation point. We
derive in \S~\ref{modflucA} the general expression of the curvature
fluctuations in the radiation era and then apply it to the case of
modulated fluctuations scenarios (\S~\ref{modflucC}).

\subsection{General Results}\label{modflucA}

Let us start by matching the inflationary stage to the radiation
era on a general constant $q$ hypersurface. From the junction
conditions (\ref{a8}), we deduce that the gravitational potential
in the radiation era is given by
\begin{equation}\label{fax1}
 \Phi_+(\eta) = \Phi_-(-\eta_*) + \frac{2-\varepsilon}{3}\left.
 \frac{{\cal H}}{q'}\right\vert_* \delta q(-\eta_*)
 \left[1-\left(\frac{\eta_*}{\eta}\right)^3\right],
\end{equation}
where $\varepsilon$ is given by Eq.~(\ref{srpara}) and depends on
the details of the inflationary stage. Using that ${\cal R}_+ =
3\Phi_+/2$, we deduce that the curvature perturbation deep in the
radiation era but long enough after the transition is given by
\begin{equation}\label{fax2}
 {\cal R}_+(\eta) = \frac{3}{2}\Phi_-(-\eta_*) + \left(1-\frac{\varepsilon}{2}\right)
 \frac{H\delta q}{\dot q}(-\eta_*).
\end{equation}
To go further, we need to specify (i) the nature of the
inflationary period which will fix $\Phi_-(\eta_-)$ and
$\varepsilon$ and (ii) the nature of the transition which will fix
$q$.

Let us note that the formula (\ref{fax2}) contains the standard case
discussed in Section~\ref{sec_infl} that is recovered by simply
choosing $q=\varphi$, taking into account that Eq.~(\ref{srpara})
implies that $H/\dot\varphi = \sqrt{4\pi/\varepsilon}/M_p$. We now
apply this general results to the case of modulated fluctuations.

\subsection{Modulated fluctuations from slow-roll hybrid inflation}\label{modflucC}

We consider the realistic scenario of a slow-roll hybrid inflationary
stage in which the value of the bifurcation point is modulated by some
light fields, as described in \S~\ref{SugraModel}. It follows that
inflation ends when $\varphi=\varphi_c(\chi_a)$ so that the parameter
$q$ introduced in \S~\ref{Sec-match} as to be chosen as
$q(\varphi,\chi_a)=\varphi-\varphi_c(\chi_a)$. Let us stress that $q$
is a function of all light fields including the inflaton. We deduce
that the quantities needed to apply the matching conditions
(\ref{a7}-\ref{a8}) are given by
\begin{equation}
 \delta q=\delta\varphi
 -\sum_a{\d\varphi_c\over\d\chi_a}\delta\chi_a
\end{equation}
and by
\begin{equation}
 \dot q =\dot\varphi
\end{equation}
because $\dot\chi_a\simeq0$. We can now apply the matching
conditions between the inflationary solution and the radiation era
solution, exactly as in Eq.~(\ref{zozo}), and making use of the
background quantities defined in Eq.~(\ref{bgd}) we obtain the
equivalent of Eq.~(\ref{fax1})
\begin{equation}
 \Phi_+(\eta)=\Phi_-(-\eta_*) + {2-\varepsilon\over 3}\left.{{\cal
 H}\over\varphi'}\right|_{*}\left[\delta\varphi(-\eta_*)-\sum_a\gamma_a\delta\chi(-\eta_*)\right]
 \left[1-\left({\eta_*\over\eta}\right)^3\right]
\end{equation}
where we have introduced the coefficients
\begin{equation}
 \gamma_a\equiv{\d\varphi_c\over\d\chi_a}(-\eta_*).
\end{equation}
Using that $\Phi_-\sim\varepsilon{\cal R}_-$, we deduce that
${\cal R}_+=3\Phi_+/2$ is given, at a late enough time
($\eta\gg\eta_*$), by
\begin{equation}
 {\cal R}_+={\cal R}_- - \left(1-{\varepsilon\over 2}\right)
\left.{H\over\dot\varphi}\right|_{*}
 \sum_a\gamma_a\,\delta\chi_a(-\eta_*).
\end{equation}
Using Eq.~(\ref{srpara}) to express $H/\dot\varphi$ at the time of
the transition, we end up with
\begin{equation}\label{61}
 {\cal R}_+={\cal R}_- - \sqrt{4\pi}{1-\varepsilon/2\over\sqrt{\varepsilon}}
 \sum_a\gamma_a\,{\delta\chi_a(-\eta_*)\over M_p}.
\end{equation}
Assuming for simplicity that there is only one light scalar field
and taking into account that ${\cal R}_-$ and $\delta\chi_a$ are
not correlated, see the discussion in \S~\ref{sec_hybrid}, we conclude that
\begin{equation}
 P_{{\cal R}_+}=P_{{\cal R}_-} + 4\pi{1-\varepsilon\over\varepsilon}
 \gamma^2{P_\chi\over M_p^2}.
\end{equation}

We now need to determine the one of $\chi_a$ that can be considered as
a test field.  Let us recall the evolution of any light field $\chi$
satisfies, in Fourier space, an equation of the form
\begin{equation}\label{c1}
 (a\delta\chi)'' +
 \left[k^2 + m^2a^2 -\frac{a''}{a}\right](a\delta\chi)=0.
\end{equation}
Assuming that the background spacetime is described by slow-roll
inflation, the bracket expression reduces to $(\nu^2-1/4)/\eta^2$
with $\nu\sim3/2 - m^2/3H^2 + \varepsilon$ if one uses
Eq.~(\ref{sr4}). The solution leads, on super-Hubble scales, to
\begin{equation}\label{c4}
 \delta\chi_k\sim\frac{H}{\sqrt{2k^3}}
 \left(-k\eta\right)^{3/2-\nu}.
\end{equation}
The power spectrum of any light field, as defined in
Eq.~(\ref{defdeP}), is thus given by
\begin{equation}\label{c5}
 P_\chi(k)= \left(\frac{H}{2\pi}\right)^2(-k\eta)^{2m^2/3H^2
 -2\varepsilon}.
\end{equation}

It follows that the power spectrum of the curvature perturbation deep
in the radiation era is given by~\footnote{We assume that the
slow-roll parameters $\varepsilon$ and $\eta$ are constant during
inflation, which is the case at first order in the slow-roll
parameters. If $\varepsilon$ had varied between the time of horizon
crossing of the modes of cosmological interest and the time of the
transition then Eq.~(\ref{PRmoins}) will be $P_{{\cal
R}_+}={1\over4\pi^2}\left({H\over M_p}\right)^2{1\over\varepsilon}
(k\eta_*)^{-2\varepsilon}\left[ (k\eta_*)^{2\eta-4\varepsilon} +
4\pi\gamma^2(1-\varepsilon_*)
{\varepsilon\over\varepsilon_*}(k\eta_*)^{2m^2/3H^2}\right]$. We will
get the same phenomenology but with $\gamma^2$ replaced by
$\gamma^2\varepsilon/\varepsilon_*$.}
\begin{equation}\label{PRmoins}
 P_{{\cal R}_+}={1\over4\pi^2}\left({H\over M_p}\right)^2{1\over\varepsilon}
 (k\eta_*)^{-2\varepsilon}\left[
 (k\eta_*)^{2\eta-4\varepsilon} +
 4\pi\gamma^2(1-\varepsilon)(k\eta_*)^{2m^2/3H^2}
 \right].
\end{equation}
Contrary to the standard case described in \S~\ref{sec_infl},
${\cal R}$ is not conserved through the transition, his jump being
given by Eq.~(\ref{61}). This is due to the fact that the
perturbation of the light fields, which were isocurvature
perturbations during inflation, are transferred to the adiabatic
mode at the beginning of the radiation era. Note also that
$\varphi'$ has been expressed in terms of the slow-roll parameters
so that $\delta\varphi$ and $\Phi_-$ combine to give ${\cal R}_+$.
Also, the standard inflationary case described in
\S~\ref{sec_infl} is recovered when $\gamma=0$, that is when the
value of the bifurcation point does not depend on any light field
and when the end of inflation takes place on a constant $\varphi$
hypersurface.

\subsection{Spectral index and the consistency relation of
Modulated fluctuations}\label{mod}

An important qualitative result we obtain here is that the modulated
fluctuations from the hybrid inflation are inevitably accompanied by
the usual inflaton fluctuations. Their relative amplitude is given by
the factor $4\pi \gamma^2(1-\epsilon)$ and their spectra are not
generically the same. While the index of the inflaton driven
fluctuations is $n_S-1=2\eta-6\epsilon$, that of the modulated
fluctuations $n_S-1=2m^2/3H^2-2\varepsilon$. The observed scalar
spectrum is therefore the sum of two power laws and the scalar
spectral index can run between these two limiting values.

Concerning gravitational waves, the super-Hubble solution in RDU
takes the form $h_+ = A_+ + {B_+}/{\eta}$ so that the matching
conditions (\ref{a8}) imply
\begin{equation}
 h_+  \sim \frac{1}{\sqrt{2k^3}}\frac{H}{M_p}(k\eta_*)^{-\varepsilon}
\end{equation}
on super-Hubble scales at any time $\eta\gg\eta_*$. It follows
that deep in the radiation era,
\begin{equation}\label{Pdeh}
 P_h(k)=
 \frac{1}{4\pi^2}\left(\frac{H}{M_p}\right)^2(k\eta_*)^{-2\varepsilon},
\end{equation}
as in the standard inflationary case.\\

In the standard case described in \S~\ref{sec_infl}, the amplitude of
the gravitational waves was set by the energy scale of inflation,
$(H/M_p)^{1/2}\sim V^{1/4}/M_p$ and their relative amplitude compared
to the scalar modes was controlled by the slow-roll parameter
$\varepsilon$ via the consistency relation~(\ref{consist}). This
implies that the detection, or limit on the amplitude, of the
gravitational waves set a constraint on the energy scale at which
inflation took place.

The modulated fluctuations lead to another effect. The consistency
relation~(\ref{consist}) becomes
\begin{equation}\label{lala}
 {T\over S}={\varepsilon\over 1 +4\pi(1-\varepsilon)\gamma^2}
\end{equation}
and the contribution of the tensor modes is always smaller in
modulated fluctuations case than in the standard inflationary
case. This can be easily understood because the modulated
fluctuations are of scalar type only and the gravitational waves
are completely insensitive to the properties of the transition, as
can be seen from the matching relation Eq.~(\ref{a8}).

The situation where $\gamma\sim{\cal O}(1)$ is particularly
interesting since $T/S$ remains of order $\varepsilon$ but a deviation
from the standard consistency relation (\ref{consist}) of order
$\varepsilon$ appears. In this regime, the scalar power spectrum is
the sum of two power laws of comparable amplitude, opening the
possibility to have a break at an observable wavelength. This in
particular the case in the explicit model presented in
\S~\ref{SugraModel}, from Eq.~(\ref{FDterm2}) we get
$\gamma^2=\sum_{a=1}^3 \gamma_a^2=2$.

When $\gamma\ge1$ that is when most the scalar perturbations are
inherited from the modulation of the transition
hypersurface the ratio $T/S$ is then much
smaller that $\varepsilon$ and we get a mechanism that damps the
gravitational waves contribution, whatever the energy scale of
inflation. On one hand, it allows the energy scale of inflation to
be higher than in the standard case and still have undetectable
gravitational waves (see also Ref.~\cite{pilo} for a scenario that
can boost the gravitational waves). In that case, gravitational
waves then appear to contribute non-significantly e.g. to the
temperature anisotropies and their detectability, see e.g.
Ref~\cite{ks1,ks2}, and strong constraints on $\varepsilon$ will
be difficult to set.

\section{Discussion and conclusion}\label{sec_concl}

 Modulated curvature fluctuations in hybrid
models can be put on firm grounds whether it is from a model
building perspective on for the computation of its
phenomenological consequences. That such models are generic might
reveal interesting at different levels. In such classes of models
indeed the metric fluctuations are not necessarily associated to
field fluctuations in the slow roll direction. It allows to break
the relations between the slow roll parameters and the shape of
the power spectrum. Analogous conclusions were reached in
Ref.~\cite{picon} where a model that decouples the spectral index
from the inflationary stage is presented. Moreover some of the
difficulties encountered in the usual hybrid models might be
circumvent. Indeed the production of topological defects at an
energy scale comparable to that of the adiabatic fluctuations, no
contribution of which have been detected, is natural in such
models forcing a tuning of the parameters subsequently causing the
suppression of ``good" scalar fluctuations (see
Ref.~\cite{Pterm}).
%First, cosmic strings are
%formed after the symmetry breaking and they contribute to
%cosmological fluctuations. As a consequence it is necessary to
%tune parameters to suppress cosmic string contribution but then
%the ``good" scalar fluctuations will be suppressed too (see
%Ref.~\cite{Pterm}).
Here we avoid this problem. Actually what we have obtained here is
a kind of decoupling of the background evolution sector form the
generation of fluctuations.
%Second, potential $V_{\rm
%sr}(\varphi)$ is detached from $V(\sigma,\varphi)$. The story of
%``usual" scalar fluctuations from the hybrid model is designed
%rather separately from the background evolution (contrary to what
%we have in chaotic model).

Are thus models arbitrary constructions or could they be
physically motivated? Let us recall that light fields, such as
moduli, are generic in e.g. string theory. They may have different
effects related to the theory of structure formation and of
interest for cosmology. They may induce a modulation of the
coupling constants that can lead (i) to a spacetime modulation of
the constants of nature, as e.g. the fine structure constant (see
Ref.~\cite{const2} for a the effect of a fluctuating light
dilaton, Ref.~\cite{const3} for an example of signature on the
cosmic microwave background and Ref.~\cite{const1} for a review),
(ii) to the generation of non-Gaussianity and (iii) to the
mechanism of modulated fluctuations described in this article.

The mechanism presented in the article shows that modulated
inflationary models correspond to a large class of models that
naturally emerges from supergravity where the couplings depending
on light moduli fields can modulate the transition to the
radiation era.
%The mechanism of {\it modulated cosmological fluctuations}
%presented in this article has shown to enlarge the standard
%inflationary predictions. In this class of models, that can be
%derived from SUSY, the couplings depend on the value of some light
%scalar field so that the transition to the radiation era is
%modulated over space. We have shown that
In such models we have more specifically shown that gravitational
waves were not affected by the modulation while scalar modes
receive an extra-contribution. The primordial power spectra
however follow a modified consistency relation, Eq.~(\ref{lala}),
which depends on a new parameter, $\gamma$, that characterizes the
dependence of the bifurcation point on the light fields. Moreover
the fact that the contribution of the gravitational waves is
always smaller in this context than in the standard inflationary
case has two consequences: (i) it allows to have a higher energy
scale for inflation and still have undetectable tensor modes but
(ii) makes the possibility of their detection via $B$-polarization
measurements more difficult. An interesting situation arises when
$\gamma\sim 1$. In that case, the tensor modes still have a
relative amplitude comparable to the standard case but (i) we have
a deviation, Eq.~(\ref{lala}), from the standard,
Eq.~(\ref{consist}), that can be hoped to be measured and (ii) the
scalar power spectrum is the sum of two power laws, opening the
door for a possible break in the range of wavelengths of interest
for cosmology.

These results should be put in parallel to other extensions of
standard inflation where the consistency relation can be modified.
Generically in multifield inflation~\cite{2field}, Eq.~(\ref{consist})
becomes $T/S=\varepsilon\sin^2\Theta$ where $\Theta$ is the
isocurvature-adiabatic correlation angle. As in our case,
$T/S<\varepsilon$. On the other hand in the curvaton
scenario~\cite{curvaton}, the density perturbations arise from the
fluctuation of a light field and primordial perturbations are entirely
from an isocurvature mode and the consistency relation becomes
$T/S\ll\varepsilon$ (see e.g. Ref.~\cite{vl}). This is analogous to a
limit in which $\gamma\gg1$.

This class of models also opens a new phenomenological avenue. Indeed
nothing prevent the fields $\chi_a$ to develop non-Gaussianity prior
to the transition as in the mechanism of Ref.~\cite{bu}. In modulated
inflation the transfer of modes is not not due to a bent of the
trajectory but to the modulation. The outcome of such a mechanism
could be the superposition of Gaussian perturbations and a
non-Gaussian ones of the same variance. Let us stress that no
non-Gaussianity is generated by the transition itself so that if
fluctuations of the light field are initially Gaussian so will be the
modulated fluctuations.

The last remarks are about the nature of modulated scalar fluctuations.
First, note that this mechanism offers an example of situation in which 
the adiabatic mode is not conserved, an issue raised in Ref.~\cite{adiab}.
Second, before the end of inflation, the moduli field fluctuations are
subdominant gravitationally and have a character of isocurvature
fluctuations.  When the equation of state is changed the moduli field
remains to be gravitationally subdominant (contrary to the curvaton
scenario where dominance of the curvaton is additionally
assumed). However, the moduli field fluctuations vary the time
hypersurface where equation of state is altered.  At this moment their
isocurvature fluctuations are transferred into adiabatic, scalar
metric fluctuations (vis the coupling which control the change of the
equation of state).  We have cosmological examples where initially
isocurvature inhomogeneities are transferred into adiabatic
fluctuations -- examples include isocurvature CDM scenario or curvaton
scenario -- in both cases the carrier of isocurvature fluctuations
become gravitationally dominant.  What is qualitatively new in the
modulated fluctuations is the fact that the carrier of fluctuations
remains always gravitationally subdominant, however, its isocurvature
fluctuations are transformed into the adiabatic one after it modulates
the jump in the equation of state.

\acknowledgments{We thank Dick Bond, Nathalie Deruelle, Christophe
Grojean, Andrei Linde, Slava Mukhanov, Marco Peloso and Filippo
Vernizzi for discussions. F.B. and J.P.U. thanks CITA for
hospitality.  The work of LK was in part was supported by NSERC
and CIAR.}

%%%%%%%%%%%%%%%%%%%%%%%%%%%%%%%%%%%%%%%%%%%%%%%%%%%%%%%%%%%%%%%%%%%%%%%

\end{document}